\include{graphicsx}

\documentclass[12pt,preprint]{aastex}

\slugcomment{Third submittal to ApJ, July 22, 2003}

\shorttitle{An Adaptive Optics Survey of M6-M7.5 Stars:}
\shortauthors{Siegler, Close, Mamajek, \& Freed}

\begin{document}


\title{An Adaptive Optics Survey of M6.0-M7.5 Stars: \\
    Discovery of Three Very Low Mass Binary Systems Including Two Probable Hyades Members}

\author{Nick Siegler, Laird M. Close, Eric E. Mamajek, \& Melanie Freed}
\affil{Steward Observatory, University of Arizona}
\email{nsiegler@as.arizona.edu}

\begin{abstract}
A survey of 30 nearby M6.0-M7.5 dwarfs with K$_s<$\,12 mag utilizing the Hokupa'a adaptive optics system at the Gemini North Telescope has discovered 3 new binary systems. All 3 systems have separations between 0.12-0.29$\arcsec$ (3\,-\,10\,AU) with similar mass ratios (q\,$>\,0.8$, $\Delta$K$_s<0.7$). This result gives further support to the suggestion that wide (a\,$>\,20$\,AU) very low mass (M$_{tot}\,<\,0.185\,$M$_\sun$) binary systems are exceedingly rare or perhaps even non-existant. The semimajor axis distribution of these systems peaks at $\sim5$\,AU, tighter than more massive M and G binary distributions which have a broad peak at separations of $\sim30$\,AU. We find a sensitivity-corrected binary fraction in the range of 5$_{-2}^{+4}\%$ for M6.0-M7.5 stars with separations a\,$>\,3$\,AU. This binary frequency is less than the $\sim32\%$ measured among early M dwarfs over the same separation range. Two of the low-mass binaries are probable Hyades open cluster members based on proper motions, cluster membership probabilities, radial velocities, and near-IR photometry. LP\,415-20 has the distinction of being the tightest (3.6\,AU) multiple system ever spatially resolved in the cluster and the companions of LP\,415-20 and LP\,475-855 are among the least massive objects ever resolved in the Hyades with estimated masses of $0.081_{-0.010}^{+0.009}$ and $0.082_{-0.009}^{+0.009}$\,M$_{\sun}$.
\end{abstract}


\section{Introduction}
High spatial resolution surveys allow for the detection of faint companions such as very low mass (VLM) stars, brown dwarfs, and possibly giant planets. These surveys also tell us something about the frequency of binary systems, their separations, mass ratios, and the variation in these parameters as a function of the primary's mass. Together they help place empirical constraints on binary star formation mechanisms. 

Many of the statistical studies on binaries have concentrated on field stars in the immediate solar vicinity. \cite{duq91} conducted a spectroscopic survey of G-dwarfs and estimated a multiplicity fraction of approximately 50$\%$ at separations greater than 3\,AU. Proceeding down the main sequence at similar separations, the fraction appears to drop monotonically: $\sim32\%$ for M0-M4 dwarfs \citep{fis92}, $\sim15\%$ for M8.0-L0.5 \citep{clo04}, $\sim10$\,-\,15$\%$ for L dwarfs \citep{bou03,giz03}, and $\sim10\%$ for T dwarfs \citep{bur03}. From the same surveys, semimajor-axis separations also appear to be a function of primary mass. While G and early M dwarfs (M0-M4) show broad separation peaks of $\sim30$\,AU, late M ($\geq$\,M8), L, and T dwarfs appear to have separations closer to $\sim2$\,-\,5\,AU. As also noted in the above references, evidence suggests that both the binary fraction {\it and} mean semimajor axis are functions of primary mass.

As part of a larger survey \citep{clo04}, we sample here a portion of the M dwarf spectral range not previously surveyed, namely M6.0-M7.5. We measure the binary frequency, binary separations, and mass ratios comparing the results to stars slightly more and less massive. The adaptive optics (AO) techniques used to make our observations and the 3 newly discovered binaries: LP\,415-20, LP\,475-855, and 2MASSW J1750129+442404 are discussed in the following section while in \S 3 we present our data reduction techniques. The two LP objects are suspected Hyades cluster members and we examine their membership probabilities in \S 4 along with the systems' derived characteristics such as distances, ages, temperatures, spectral types, and masses. We conclude by discussing the binary frequency and separation distribution of M6.0-M7.5 dwarfs in \S 5.

\section{An Adaptive Optics Survey of Nearby M Stars}
\subsection{The Survey} \label{bozomath}
In 2001, we commenced the first ground-based, AO infrared survey searching for faint companions (giant planets, brown dwarfs, and VLM\footnote{In this paper we refer to VLM binaries as those with M$_{tot}$\,$<$\,0.185\,M$_{\sun}$.} stars) around VLM stars at the Gemini North telescope. The survey strategy combines 4 observational advantages: 1) the low luminosity of VLM stars, 2) the near diffraction-limited resolution ($\sim0.1\arcsec$) of AO in the near-IR (NIR; 1-2.5 $\mu$m), 3) high sensitivity to cool objects in the NIR, 4) a large aperture telescope. 

The overall survey targets M6.0-L0.5 dwarf objects with K$_s<12$ from mainly 2MASS stars listed in Cruz et al. 2003 (in preparation), \cite{rei02}, \cite{giz00}, and \cite{kir00}. To date, we have observed 69 of these objects detecting 12 binary systems, 10 of which are newly discovered. In \cite{clo03} we presented a summary of our initial findings targetting M8.0-M9.5 stars: 20 objects observed, 4 binary systems discovered, with at least one of the four companions most likely a brown dwarf with spectral type $\sim$\,L7. \cite{fre02} presented the tightest (3\,AU) brown dwarf ($\sim$\,L7.5) ever imaged around a star, LHS 2397a. 
\cite{clo04} summarize the whole survey and discuss implications for brown dwarf/VLM binary formation theory. Here we present, in detail, the results from a subset of our survey targetting 30 M6.0-M7.5 stars. The selection constraints of our observed objects, mainly from \cite{giz00}, are galactic latitude b $>20\degr$ (for 0\,$<\,\alpha\,<$\,4.5 hours, $\delta\,<\,30\degr$) and K$_s$\,$<$\,12.

\subsection{Guiding on Very Faint VLM Guide Stars} \label{bozomath}
One of the challenges when utilizing AO is locating sufficiently bright guide stars close enough to one's scientific object such that the isoplanatic error does not reduce the Strehl ratio to unusable levels. When observing $20\degr$ out of the galactic plane, as we are here, the challenge becomes even more significant since there are even fewer bright natural guide stars (V$<$16). Thus the capability of guiding an AO system directly on one's target object in the absence of bright natural or artificial guide stars becomes an important scientific capability. However, if the science object itself is a very faint VLM star (V$\simeq19$, I\,=\,14\,-\,16), locking an AO system may be quite difficult. Most AO systems cannot guide on such faint targets. As explained elsewhere \citep{clo02,sie02}, working exceptions are curvature-based AO systems which employ red-sensitive, photon-counting, avalanche photodiodes (APDs) with effectively zero read-noise in their wavefront sensors. We utilized the recently decomissioned University of Hawaii curvature AO system Hokupa'a \citep{gra98} which was a visitor instrument on the 8\,m Gemini North Telescope. Hokupa'a is a highly sensitive curvature AO system and well suited to guiding on nearby, faint, late-type M stars producing $0.13\arcsec$ images at K$^\prime$ band. This capability has allowed us to survey the nearest late M and early L stars in order to characterize their binary nature and search for VLM stellar and substellar companions. Two of the discovery binaries in this paper, LP\,415-20 and LP\,475-855, both had V magnitudes of 19.2 \citep[respectively]{bry94,leg94}.

\subsection{Observations} \label{bozomath}
The M6.0-M7.5 survey was conducted during 4 observing runs in 2001 and 2002 at the Gemini North telescope on Mauna Kea. A total of 30 objects were observed and 3 new binary systems were discovered. Table \ref{tbl-1} lists the objects observed with no likely physical companion detections between 0.1$\arcsec$\,-\,15$\arcsec$ ($\sim3$\,-\,300\,AU). Table \ref{tbl-2} lists the 3 newly discovered binary systems, their photometry, and the dates observed.

Each of the observations were made by dithering over 4 different positions on the University of Hawaii's Quick IR Camera (QUIRC) containing a 1024$\times$1024 infrared array detector with a platescale of 0.0199\arcsec/ pixel \citep{hod96}. At each dither position we took 3$\times$10\,s exposures at J, H, K$^{\prime}$ bands and longer 3$\times$30\,s exposures at H to improve the detection sensitivity of faint companions. Each object was additionally analyzed within its halo for even nearer, very faint companion objects as explained in the following section. 

\placetable{tbl-1}
\placetable{tbl-2}

\section{Reduction}

The images were reduced using an AO data reduction pipeline written in the IRAF language as described in \cite{clo02}. Unlike conventional NIR photometry, we purposefully disabled the Cassegrain rotator so that the pupil image remained fixed with respect to the camera and detector. The residual PSF aberrations including telescope primary and secondary structures remained static and fixed in the final images, allowing for comparative identification of companions without confusion from the rotation of PSF artifacts. The pipeline compensates for this in the final images by rotating each image by the parallactic angle as described in \cite{clo04}.

\begin{figure}
\includegraphics[angle=0,width=\columnwidth]{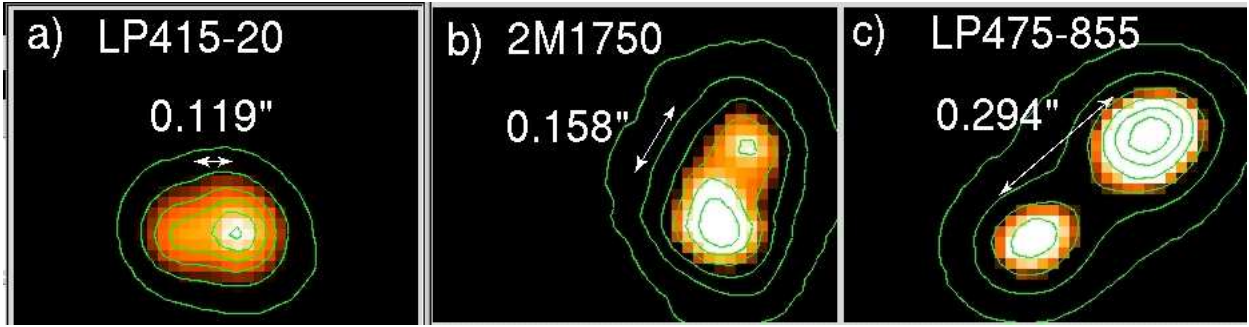}
\caption{The three newly discovered binary systems observed at a resolution of $0.13\arcsec$ in K$^\prime$ (a) 9$\times$10\,s image of the LP\,415-20 binary observed on 2002 February 07, UT. (b) 1$\times$10\,s image of the 2MASSW J1750129+442404 binary system observed on 2002 April 25, UT. (c) 9$\times$10\,s image of the LP\,475-855 binary observed on 2001 September 22, UT. Platescale is 0.0199$\arcsec$/pixel. The contours are linear at the 90, 75, 60, 45, 25, and 15\% levels. North is up and east is to the left.
\label{fig1}} 
\end{figure}

The pipeline produces final unsaturated 120\,s exposures in three NIR bandpasses: J (FWHM $\simeq0.15\arcsec$), H (FWHM $\simeq0.14\arcsec$), and K$^{\prime}$ (FWHM $\simeq0.13 \arcsec$) with deep 360\,s exposures (FWHM $\simeq0.14 \arcsec$) at H band for each observed binary system. The dithering of the shorter exposures produces a final 30$\arcsec\times30\arcsec$ image with the most sensitive overlap region (10$\arcsec\times10\arcsec$) centered on the binary. A technique that we used to search for tight companions potentially orbiting within the halo of the primary involved Fourier filtering the low spatial frequency components leaving behind only high frequency residuals in the PSF. With the Cassegrain rotator disengaged, potential companions within the halo could have been easily differentiated from PSF artifacts. However, no very faint companions were found in the halo of our primaries using this technique. All 3 binary systems were detected from reductions of the shorter exposures. Figure \ref{fig1} shows K$^{\prime}$ images of each of the new systems.

\placefigure{fig1} 

Photometry was performed using the DAOPHOT PSF fitting photometry package in IRAF. The PSFs used were reduced 12$\times$10\,s unsaturated single stars observed during the same night with similar IR brightness, spectral type, and air mass. Adjustments to the PSF were made in some cases to match slight PSF elongations due to the AO system. This technique was successfully employed for the data reduction of the 0.29$\arcsec$ separated LP\,475-855 and the K$^{\prime}$ images of each of the 2 tightly separated LP systems. The errors in $\Delta$mag are the differences in the photometry between 2 similar PSF stars. 

DAOPHOT could not successfully separate the two blended components of LP\,415-20 and 2M1750 in the J and H bands due to the lower Strehl ratios of the AO system at these shorter wavelengths. Instead, we use a technique applied by \cite{fre02} which uses the annulus photometry task {\it phot} on the images that have had their low spatial frequencies previously removed. This purely differential technique preserves the relative magnitude difference while removing sufficient primary halo flux to reveal the companions clearly seen in K$^{\prime}$. The technique gave reliable $\Delta$mags and was verified on binary images with known $\Delta$mags with, however, increased photometric uncertainties of about 0.10 mag.

We calculate individual fluxes and their uncertainties from the measured flux ratios of the binaries and the integrated 2MASS apparent magnitudes \citep{giz00}, along with their respective uncertainties. In order to work with the 2MASS blended K$_s$ photometry we seek a conversion from $\Delta$K$^\prime$ to $\Delta$K$_s$. We find $\Delta$K$^\prime\simeq\Delta$K$_s$ to within 0.02 mag for $0.8<\,$(J-K$_s$)$\,<\,$2.0 according to the models of \cite{cha00} for objects between 0.05-0.10\,M$_{\sun}$. Hence, we simply assume $\Delta$K$^\prime$\,=\,$\Delta$K$_s$. Table 3 lists the photometry and derived characteristics of the new binary systems. Overall uncertainty in the flux includes this $\Delta$mag error along with the 2MASS error for the integrated image per bandpass.

The platescale and orientation of the IR camera were determined from a short exposure of the Trapezium cluster in Orion and compared to published positions \citep{sim99}. From these observations a platescale of $ 0.0199\pm 0.0002\arcsec$/pix and an orientation of the Y-axis ($0.3\pm 0.3$ degrees E of north) was determined. 

\placetable{tbl-3}

\section{Analysis}
\subsection{Are the Companions Physically Related to the Primaries?} \label{bozomath}
 Our target list consists mainly of objects $>20\degr$ above the galactic plane so we did not expect to observe many background stars in our images. From the 69 objects already observed in the total survey ($6.2\times10^4$ square arcsec), we did not detect any additional (J\,-\,K$_s)>0.8$ mag (background) objects in any of the fields. Thus the probability of a chance projection of an object within 0.5$\arcsec$ of the primary is estimated to be $<1.3\times 10^{-5}$. Furthermore, if we consider the possibility that our apparent companions are actually background M6-M8 dwarfs fainter by a $\Delta$ magnitude of 1.5 mag (ie. they are twice as distant as their primaries), we estimate the probability of a background M6-M8 (with a stellar density of 0.007/pc$^3$; \cite{rei97}) appearing within $0.5\arcsec$ of any of our 30 targets to be $\sim3\times10^{-7}$. Additionally, none of the companion images appear spatially extended as might be expected for background galaxies. Therefore, we conclude that all three very red objects are physically associated with their primaries and hereafter will refer to them as LP\,415-20B, LP\,475-855B, and 2MASSW\,J1750129+442404B (2M1750B, hereafter).

\subsection{Distances} \label{bozomath}
\begin{figure}
\includegraphics[angle=90,width=\columnwidth]{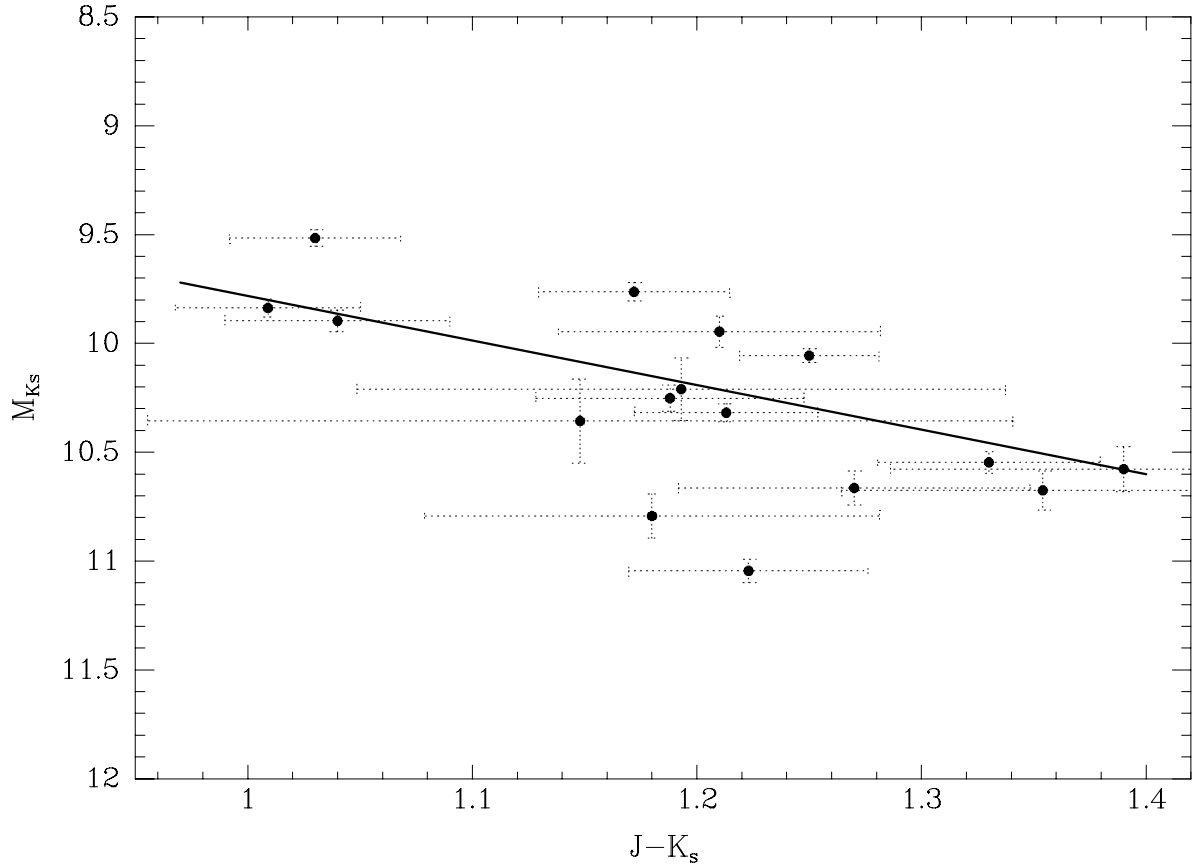}
\caption{Derived M$_{K_s}$ magnitudes versus J-K$_s$ colors using only 2MASS measured K$_s$ photometry and \cite{dah02} triginometric parallaxes. Figure valid for M6.5-L1.0 dwarfs with the linear fit M$_{K_s}$ = 7.65 + 2.13(J-K$_s$) shown. The scatter about this fit is $\sigma$\,=\,0.33 mag.  
\label{fig6}} 
\end{figure}

With none of the 3 binary systems having published trigonometric parallaxes, we estimate photometric distances based on trigonometric parallaxes of other nearby VLM stars \citep{dah02}. We construct a color-magnitude diagram (CMD) of M6.5-L1 stars in Figure \ref{fig6} using 2MASS photometry and parallaxes from \cite{dah02}. We cut the fit at M6.5 based on evidence in the Dahn/2MASS photometry of an upward break just blueward of J-K$_s\approx0.95$. \cite{giz00} show similar evidence in their Figure 1 for Hyades members as do \cite{leg94}. In Figure \ref{fig4} we overplot our CMD from Figure \ref{fig6} onto a CMD of the faintest candidate low mass Hyads from \cite{leg94}. The \cite{dah02}/2MASS data match very well with the \cite{leg94} Hyades results redward of $\sim0.95$\,mag. With all our binary components having J-K$_s\gtrsim0.97$, we believe that the linear fit of Figure \ref{fig6} redward of J-K$_s\approx0.95$ to be justified for either field or apparent Hyades members as well. This also suggests that the differences in the CMD for the Hyades and other field stars due to astrophysical effects (age, metallicity, log(g), etc.) are less than our observational errors. We find (Fig. \ref{fig6}) a linear least-squares fit of M$_{K_s} = 7.65 + 2.13\,$(J-K$_s)$ for the spectral range M6.5-L1. The rms scatter in M$_{K_s}$ is 0.33 mag which when taken in quadrature with the uncertainty in the photometry gives the overall M$_{K_s}$ error. The result is nearly identical to the CMD fit of \cite{giz00} using parallaxes of late M dwarfs. We then use the distance modulus of the primary to estimate the distances to the binaries. This method estimates photometric distances of $30\pm5$, $29\pm5$, and $31\pm6$ pc for LP\,415-20A, LP\,475-855A, and 2M1750A, respectively (Table \ref{tbl-3}).

\bigskip

Before we commence an analysis of the ages, temperatures, spectral types, and masses of these new objects, we consider the possibility that 2 of the discovered binaries are members of the Hyades open cluster. Assigning cluster membership is quite valuable in that it provides credible ages and abundances. In turn, known ages place tighter constraints on estimated temperatures and masses using theoretical evolutionary tracks. LP\,415-20 and LP\,475-855 are only offset by approximately 3$\degr$ from the cluster center which is well within the $\sim12\degr$ tidal radius of the cluster. Furthermore, 95$\%$ of 190 cluster members from a Hipparcos study of the Hyades \citep{per98} have trigonometric distances between 25 and 73\,pc. LP\,415-20 and LP\,475-855 have photometric distances of $30\pm5$ and $29\pm5$\,pc, respectively.  We examine further the membership probabilities of these two objects in the next section.

\subsection{LP\,415-20 and LP\,475-855: Hyades Members?}\label{bozomath} 
Two of our objects, LP\,415-20 and LP\,475-855, have been previously discussed in the literature as candidate Hyades members (LP\,415-20\,=\,Bryja\,262; \cite{bry92,egg93,rei93,bry94,sta94,giz00}; LP\,475-855\,=\,[LHD]\,042614+13312\,=\,2MASSW\,J0429028+133759; \cite{leg89,egg93,leg94,giz00}) with varying conclusions over their membership. Classifying individual objects as {\it bona fide} cluster members is often problematic, especially in a nearby open cluster such as the Hyades whose proximity (mean distance $\sim45$\,pc) means that it occupies more than 200 square degrees of sky. Besides the proximity in the sky to the cluster center and consistent distances, we consider several additional factors in evaluating if these two objects are members of the Hyades: 1) proper motions, 2) cluster membership probabilities, 3) radial velocities, and 4) NIR photometry. 

When we compare the proper motions and radial velocities of LP\,415-20 and LP\,475-855 to that of the Hyades, we find it useful to have a kinematic model for field stars to which we can compare them. We generate a Monte Carlo simulation of 100,000 synthetic stars with the kinematic properties of ultracool M dwarfs (SpT\,$\geq$\,M7) at the positions and photometric distances ($\sim$30 pc) of the 2 candidate systems, LP\,415-20 and LP\,475-855. For our kinematic model, we use the velocity ellipsoid for ultracool M dwarfs given by Reid et al.\,(2002; eq.\,2). This defines a Local Standard of Rest (LSR) for ultracool dwarfs as well as a 3D velocity dispersion. 

The likelihood of membership is best judged when all factors are evaluated collectively. We discuss each of them and list both measured data from the literature and predicted quantities for Hyades members in Table \ref{tbl-4}.

\subsubsection{Proper Motions}\label{bozomath}
We show in Figure \ref{fig2} a proper motion vector diagram showing the proper motions of both candidate objects \citep{giz00} along with the proper motions of Tycho-2 Catalogue stars \citep{hog02} within 5$\degr$. The outlined sectors represent the regions where the proper motions of Hyades cluster members are expected to lie according to \cite{bry94} - between 90$\degr$ and 135$\degr$ east of north and proper motions in the range 74-140 mas/yr. The proper motions of LP\,415-20 and LP\,475-855 fall within this proper motion space while the majority of field stars covers a distinctively different portion of proper motion space.

What do the proper motions of field late-M stars look like with respect to the positions of the 2 LP objects in Figure \ref{fig2}? The proper motion of the mean space motion (LSR) of ultracool dwarfs as a function of distance is shown as a diagonal line (solar reflex motion). The centroid of the proper motion distribution of 100,000 ultracool M dwarfs from our simulation (large open circles) is the LSR motion at the photometric distances for LP\,415-20 and LP\,475-855. Hence, the LSR is located in a distinctively different portion of proper motion space from the LP stars and Hyades loci. In the next section, we will attempt to quantify the probability of a field star moving at this solar reflex motion is a member of the Hyades using the moving cluster method.

\subsubsection{Cluster Membership Probabilities}\label{bozomath}
Convergent point analysis selects cluster members based on proper motions and positions without any assumptions of their spatial distribution. Therefore, the technique can identify moving groups which share a constant velocity vector. However, a star moving towards a cluster convergent point does not prove cluster membership but a non-converging motion can exclude it. We transform the equatorial proper motion components ($\mu_\alpha, \mu_\delta$) for our objects to the proper motion towards the Hyades convergent point ($\mu_\parallel$) and perpendicular to the convergent point-star great circle ($\mu_\perp$). The orthogonal component $\mu_\perp$ can be considered a rough indicator of the relative probability of cluster membership whereby the larger the absolute value of $\mu_\perp$, the lower the likelihood of membership. To estimate membership probabilities, we use the formulae from \citet[eq.~23 and those in $\S2.1$]{deb99}, the \cite{deb01} Hyades convergent point solution $(\alpha_{cp}$,$\delta_{cp})$\,=\,($97.29\degr, +6.86\degr)$, S$_{tot}$\,=\,46.38 km/s, v$_{disp}$\,=\,0.3 km/s, and proper motion uncertainties of 10\,mas/yr in $\alpha$ and $\delta$ for the 2 LP objects (J. Gizis, private communication). We calculate membership probabilities for LP\,415-20 and LP\,475-855 of 90$\%$ and 98$\%$, respectively. \cite{rei93} assigned LP\,415-20 a 15-50$\%$ membership probability but using a pre-Hipparcos convergent point solution for the Hyades.

Quantifying the discussion from the previous section, hypothetical field stars at the positions of LP\,415-20 and LP\,475-855  at the photometric distance ($\sim30$\,pc) moving with the LSR for ultracool dwarfs would have Hyades membership probabilities of well under 1$\%$ (10$^{-8}$). We find that only 2.5$\%$ and 2.4$\%$ of the Monte Carlo sample of ultracool M dwarf field stars have membership probabilities higher than 90$\%$, for LP\,415-20 and LP\,475-855, respectively.

\subsubsection{Radial Velocities}\label{bozomath}
The moving cluster method also predicts radial velocities for cluster members. Using the positions of our binary systems and the Hyades convergent point solution, the method predicts $38.7\pm0.3$ km/s and $40.1\pm0.3$ km/s for LP\,415-20 and LP\,475-855, respectively. The uncertainty in radial velocities is due to just the internal velocity dispersion following \cite{deb01}. LP\,415-20's radial velocity has been observed at $43.9\pm3.9$ km/s \citep{sta94} and $36\pm4$ km/s \citep{jon96}. \cite{rei02} measured $44.3\pm2.0$ km/s for LP\,475-855. The values are not inconsistent with cluster membership, especially if orbital motions are considered (the radial velocity amplitude of the primary is of order 2\,-\,3\,km/s using the masses we derive in \S 4.5).

Figure \ref{fig3} shows a predicted radial velocity distribution along the line of sight of the Hyades for an M-dwarf sample as calculated by \cite{sta94}. Within this distribution, \cite{sta94} find a narrow range of radial velocities for known low mass Hyades. The measured radial velocities of the two candidate objects fall in this range of radial velocities for Hyades members. 

Quantifying this further we ask: What fraction of field stars have radial velocities within a given threshold of the predicted Hyades velocity? Our Monte Carlo simulation shows that only 12$\%$ of ultracool M-dwarf field stars would have radial velocities within $\pm5$\,km/s of the Hyades mean. The uncertainty in our predicted radial velocities are mostly in the internal velocity dispersion of the cluster and is relatively insensitive to variations in the assumed distances, the observational radial velocity errors (2-4\,km/s), and the proper motion uncertainties. Our simulation, using a velocity ellipsoid for ultracool dwarfs, predicts radial velocities of +25\,km/s for stars moving with the LSR at the positions of both LP objects .

\subsubsection{NIR Photometry}\label{bozomath}
\begin{figure}
\includegraphics[angle=90,width=\columnwidth]{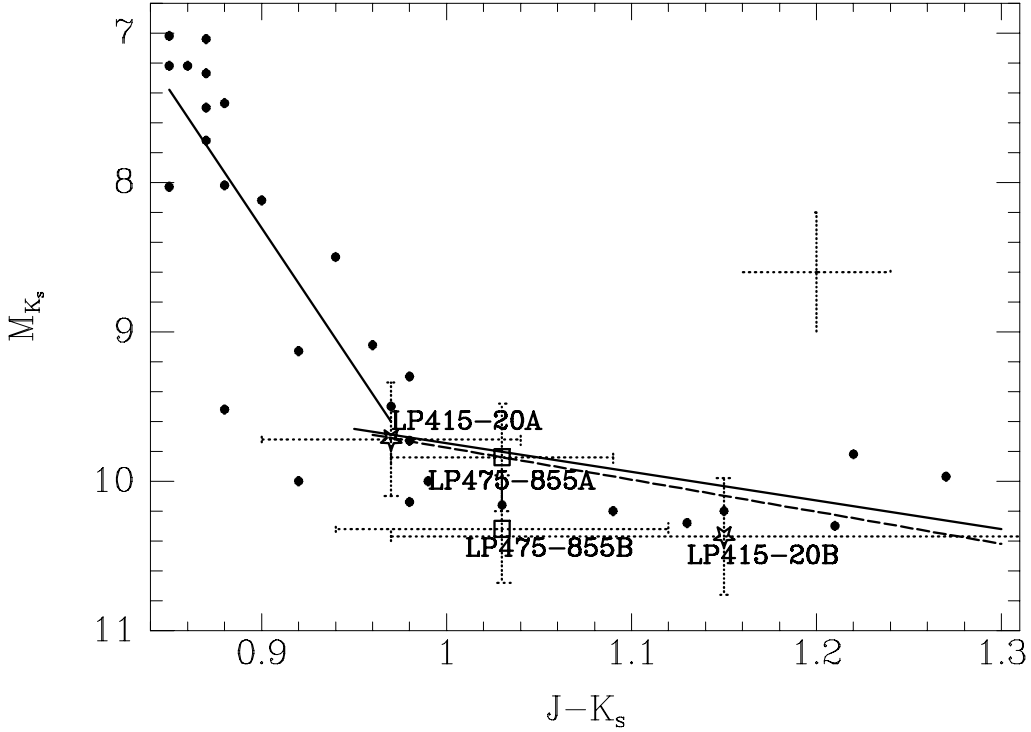}
\caption{CMD showing the two candidate Hyades systems (open stars: LP\,415-20, open squares: LP\,475-855) at photometric derived distances (30 and 39\,pc, respectively) overplotted onto a CMD of the faintest candidate low mass Hyads from \cite{leg94}. We convert, for the purposes of this figure, K$_{CIT}$ to (K$_s)_{2MASS}$ by subtracting 0.02 mag \citep{car01}. We show a typical error bar in the upper right for the \cite{leg94} data. We fit the data as two distinct regions: a ``blueward'' segment fits the steeply rising stars blueward of J-K$_s$\,=\,1.0 ($\sigma=0.59$) and a ``redward'' segment fits objects with M$_{K_s}$\,$>$\,9 (bold lines). The dashed line is the best fit of M6.5-L1.0 field stars using 2MASS photometry and trigonometric distances from Figure \ref{fig6} ($\sigma=0.33)$. There is good agreement between the two lines.
\label{fig4}} 
\end{figure}

\begin{figure}
\includegraphics[angle=0,width=\columnwidth]{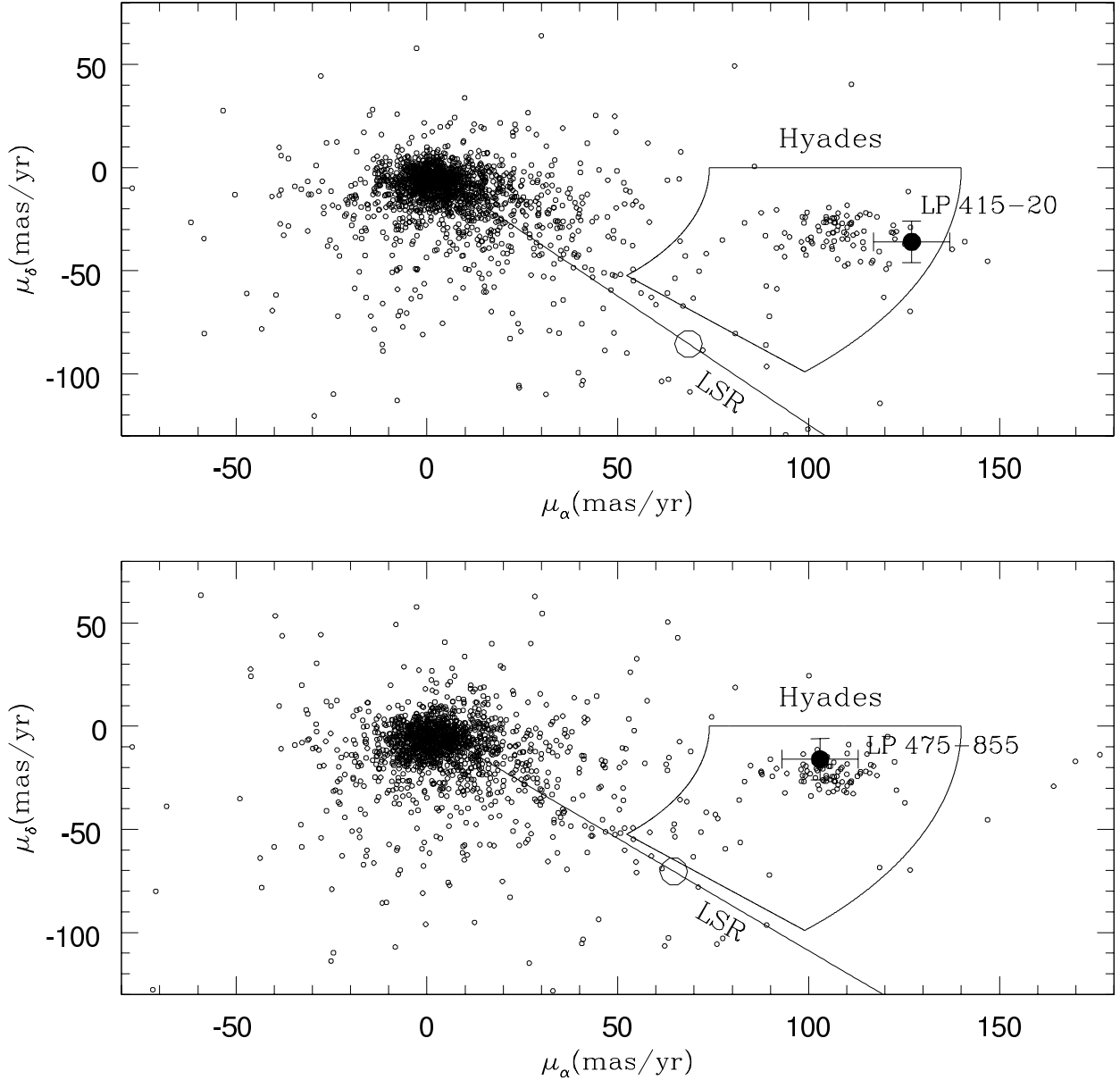}
\caption{Proper motion diagram for LP\,415-20 (top) and LP\,475-855 (bottom). 
The measured proper motions for these stars \citep{giz00} are indicated by filled circles with error bars. For reference only, we plot the Hyades cluster member selection sectors as defined by \cite{bry94}. The diagonal line defines the Local Standard of Rest (LSR) for ultracool M dwarfs \citep{rei02}. The large open circle on the line is the LSR motion at the photometric distances predicted for the LP stars ($\sim30$\,pc). The small open circles are the proper motions for all Tycho-2 stars within 5$\degr$ of LP\,415-20 and LP\,475-855 (N=1856 and 1679 stars, respectively). 
}
\label{fig2} 
\end{figure}

\begin{figure}
\includegraphics[angle=0,width=\columnwidth]{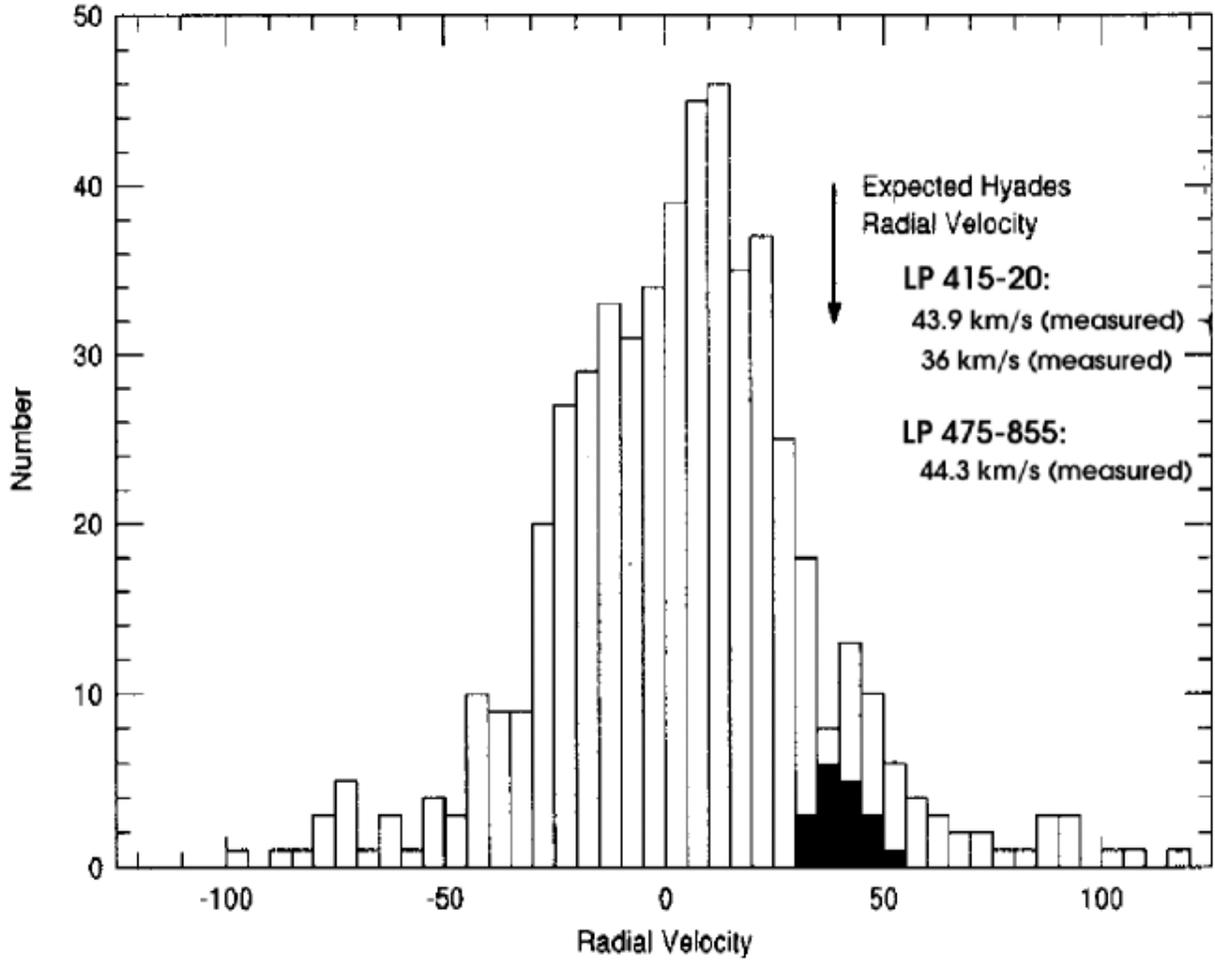}
\caption{Radial velocity distribution from \cite{sta94} for field M dwarfs along the line of sight to the Hyades plotted along with 18 observed radial velocities of known Hyades objects (dark region). The measured radial velocities of our two candidate Hyades objects fall within this colored region (LP\,415-20: \cite{sta94,jon96}, LP\,475-855: \cite{rei02}). 
\label{fig3}} 
\end{figure}

\cite{rei00}, using a Hyades moving cluster distance ($\sim40$\,pc), show the unresolved system LP\,415-20 to lie 0.5\,-\,1.0 mag above the main sequence for late-type Hyades dwarfs on a [M$_V$,\,(V-I)] CMD. This is consistent with an over-luminous, unresolved binary. Conducting a photometric study of low mass stars in the Hyades, \cite{leg94} considered LP\,475-855 too luminous to be a Hyades member according to its position on a [M$_K$,\,(V-K)] CMD using moving cluster derived distances \citep{rei93}. In Figure \ref{fig4} we plot the positions of our two candidates' A and B components at their photometric distances on a [M$_K$,\,(J-K)] CMD using NIR photometry from \cite{leg94}. Both companions are within 1\,$\sigma$ of the average Hyades main sequence.

\bigskip

We note that LP\,415-20 and LP\,475-855 are along the line of sight towards the Taurus-Auriga star forming region which is $\sim150$\,pc away. We discount membership to the distant Taurus clouds since their proper motions are too large. The average proper motion of stars in the Taurus-Auriga molecular clouds is ($\mu_\alpha$, $\mu_\delta$) = (+4, -17) mas/yr \citep{fri97} while LP\,415-20 and LP\,475-855 have proper motions of (+127, -36) mas/yr and (+103, -16) mas/yr, respectively \citep{giz00}. Not only are the proper motions nearly an order of magnitude too high, but their velocity direction is nearly 90$\degr$ off. Also, placing the LP stars at d\,=\,150\,pc leads to V$_{tan}$\,=\,94 km/s and 104 km/s, respectively. Tangential velocities this high are not observed for young stars.

It should be noted that the moving cluster method makes its own distance predictions of 41$_{-6}^{+8}$ pc and 47$_{-6}^{+8}$ pc for LP\,415-20 and LP\,475-855, respectively (using \cite{deb01} convergent point solution and \cite{giz00} proper motions; the errors are dominated by the uncertainties in proper motions of the primaries and the internal velocity dispersion of the cluster). The greater distances are both within $2\,\sigma$ of the photometric-derived distances and may be due to a potential underestimation of the photometric distances in the M$_{K_s}$ upturn shown in Figure \ref{fig4}. While these larger distances are statistically consistent with the photometric distances, they correspond to more luminous primaries (M$_{K_s}$\,=\,9.1 and 8.8 mag for LP\,415-20 and LP\,475-855, respectively). As we shall examine in \S 4.5, these absolute magnitudes will correspond to stars with hotter temperatures \citep{cha00} and estimated spectral types of $\sim$M5 and M4 \citep{dah02}. Our measured $\Delta$K$_s$ values would therefore imply companion magnitudes of M$_{K_s}$\,=\,9.8 and 9.3\,mag which correspond to predicted spectral types of M7.5 and M6.0, respectively. Consequently, we would expect blended spectra to be approximately M5.0-M6.0 for the LP binaries. However, the LP objects have measured blended spectra of M7\,$\pm1.0$ and M7.5\,$\pm1.0$ spectral types (LP\,415-20: \cite{bry94,giz00}; LP\,475-855: \cite{giz00,rei02}). Alternatively, the photometric distances predict temperatures and spectral types, as we shall examine, more consistent with spectral observation and hence we will use the photometric distances throughout the rest of this paper. 

In summary, based on photometric distances, proper motions, moving cluster probabilities, radial velocities, and NIR photometry, LP\,415-20 and LP\,475-855 appear to have characteristics consistent with membership in the Hyades open cluster. Both LP\,415-20 and LP\,475-855 have radial velocities within 5\,km/s of the Hyades mean radial velocity and Hyades membership probabilities of $\geq90\%$. Our Monte Carlo simulations show that the fraction of field stars with Hyades membership probabilities $\geq90\%$ {\it and} radial velocities within 5\,km/s of the Hyades mean are a mere $\sim0.3\%$. Trigonometric parallaxes of the two LP objects, however, would be the decisive piece of data which would enable definitive membership classification. Table 4 summarizes these results and compares them to values near the cluster's center.

\subsection{Ages} \label{bozomath}
Membership of LP\,415-20 and LP\,475-855 in the Hyades open cluster provides a narrow age range of $625\pm50$ Myr for these objects \citep{per98}. In addition, high resolution spectra taken of LP\,475-855 \citep{rei02} found no Li, inferring that the primary is at least 0.1\,Gyr; the other two objects have not yet been analyzed for Li. Determining an age estimate for 2M1750 is a difficult task with no Li detections to provide an upper limit. In this case we conservatively assume a typical range of common ages in the solar neighborhood of 0.6\,-\,7.5 Gyr \citep{cal99} and estimate a mean age of $\sim3$ Gyr considering the binary's tangential velocity (V$_{tan}=17$ km/s) is on the lower end of the tangential velocities listed in \cite{giz00}.

\subsection{Spectral Types and Temperatures} \label{bozomath}
\begin{figure}
\includegraphics[angle=90,width=\columnwidth]{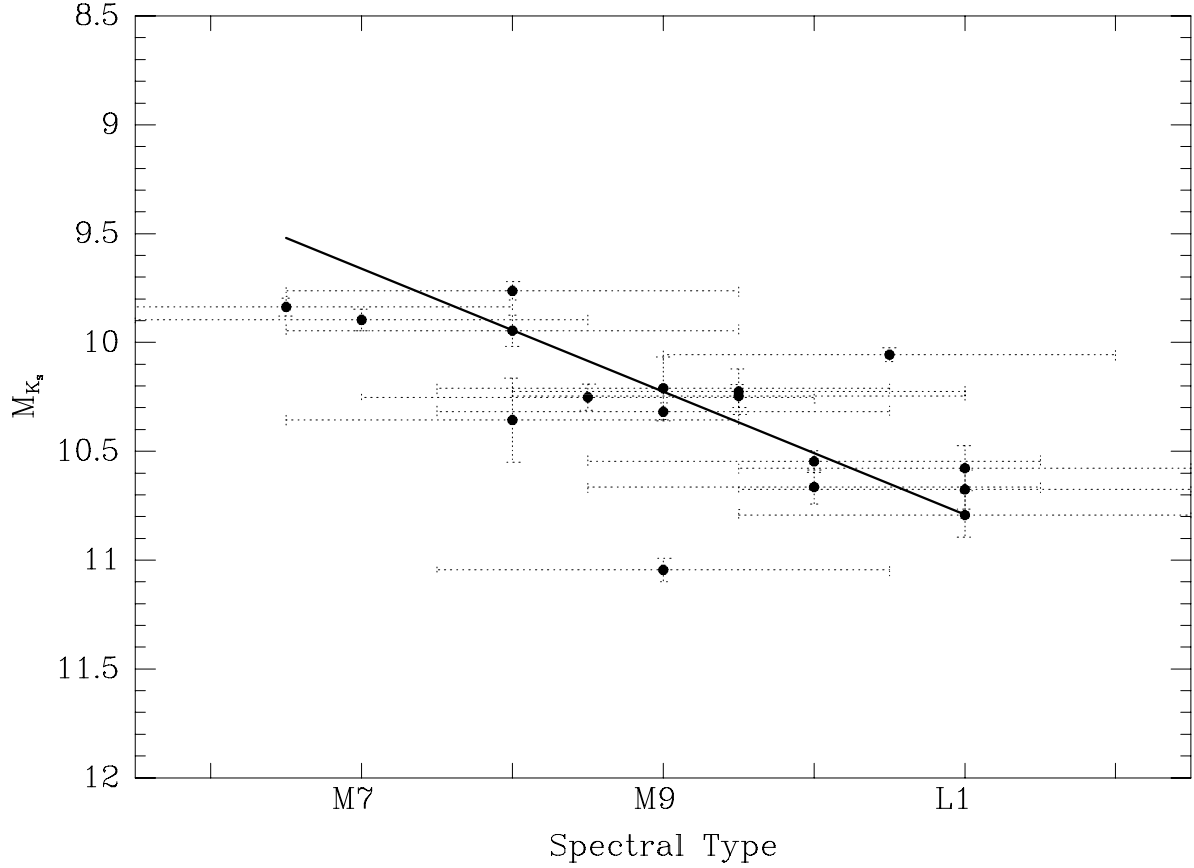}
\caption{Derived spectral types for absolute K$_s$ magnitudes. Data obtained from \cite{dah02} with corresponding photometry from 2MASS. Linear fit of spectral type is SpT\,=\,3.54M$_{K_s}$\,-\,27.20, valid for M6.5\,-\,L1.0 dwarfs (a value of 10 is defined as an L0). The scatter about this fit is $\sigma_{RMS}$\,=\,0.85 spectral types and when taken in quadrature with the M$_{K_s}$ uncertainty gives an uncertainty of 1.5 spectral types.
\label{fig10}} 
\end{figure}

While we do not have spatially resolved spectra of the individual components in any of the 3 new systems, LP\,415-20 and LP\,475-855 have had blended spectra taken by several groups confirming their M dwarf classification. \cite{bry94} concluded LP\,415-20 was an M6.5 while \cite{giz00} estimated it as M7.5. \cite{giz00} classified LP\,475-855  (M7.0) and 2M1750 (M7.5). We estimate the spectral types of each of the binary components by using the available trigonometric parallaxes for M6.5-L1.0 dwarfs \citep{dah02} that have corresponding released 2MASS photometry and fit them to their known spectral types \citep{dah02}. We find (Fig. \ref{fig10}) a linear least-squares fit of SpT = 4.54M$_{K_s}$ - 27.20 where the spectral types are quantified (eg. SpT\,=\,8 is an M8, SpT\,=\,10 is an L0, etc). The residual scatter is 0.85 spectral types which when taken in quadrature with the uncertainty in M$_{K_s}$ gives an overall uncertainty of 1.5 spectral types. The results are listed in column 6 of Table \ref{tbl-3}. Our results are consistent with the past classifications from the literature. 

Effective temperatures of the binary components are estimated from the DUSTY evolutionary tracks using calculated M$_{K_s}$ values and estimated ages (Figures 7-9). The results are listed in Table \ref{tbl-5}. We compare in Table 5 our results with those predicted for M6.5-L1.0 field stars from the \cite{dah02} spectral type - T$_{eff}$ relationship. There is very good agreement between the two results.

\subsection{Masses} \label{bozomath}
We also utilize the DUSTY evolutionary tracks \citep{cha00} to estimate the component masses of the three binary systems (see Figures 7-9). The tracks provide theoretical estimates for both stellar and substellar masses as a function of both absolute K$_s$ magnitude and age. The tracks are calibrated for the K$_s$ bandpass (I. Baraffe, private communication) and we slightly extrapolate the isochrones from 0.10 to 0.11\,M$_\sun$ to estimate the upper mass limits of the primaries. A primary's mass is estimated knowing both its age and absolute K$_s$. The companion's absolute magnitude is simply determined by adding the measured $\Delta$K$_s$ to its primary's M$_{K_s}$. The crosses indicate the best estimates of where the binary components lie on the tracks and their uncertainties are represented by the shaded regions. The primary's region of uncertainty is indicated in bold outline and the companion's is dashed. Note that their uncertainty regions overlap. The maximum mass is related to the minimum M$_{K_s}$ at the oldest possible age; the minimum mass is related to the maximum M$_{K_s}$ at the youngest possible age.  The uncertainty in the masses is largely due to both the uncertainty in the J band photometry, where the AO's resolution was poorest, and the error ($\sigma$\,=\,0.33 mag) in our CMD linear fit from Figure \ref{fig6}. The uncertainty in J band photometry propagates as error in the J-K$_s$ color and hence in the M$_{K_s}$ determination. Table \ref{tbl-3} lists the estimated masses for all 3 binary systems. All 3 systems' primary masses are consistent with M7-type dwarfs and their secondaries all appear stellar, however, their uncertainties extend into the substellar region. The extended regions of uncertainty reinforce the need for trigonometric parallaxes of these objects.

LP\,415-20B and LP\,475-855B are among the lowest mass objects spatially resolved to date in the Hyades with best estimate masses of 0.081$_{-0.010}^{+0.009}$\,M$_{\sun}$ and 0.082$_{-0.009}^{+0.009}$\,M$_{\sun}$, respectively (there is a $\sim$\,0.25\,mas spectroscopic brown dwarf candidate companion RHy 403B discussed in \cite{rei00} and a 0.04\,M$_{\sun}$ brown dwarf candidate companion of V471 Tauri \citep{gui01}; LH\,0418+13 is estimated at 0.083\,M$_{\sun}$ in \cite{rei99}). Additionally, the LP\,415-20 system would be the tightest system resolved to date in the cluster with a projected separation of $3.6\pm0.7$\,AU. The tightest known binaries in the Hyades are the 2 confirmed spectroscopic binaries from \cite{rei00}, RHy\,42 and RHy\,403, with semimajor axis of 0.61 and 0.0047\,AU, respectively. The tightest resolved binary in the Hyades was previously RHy\,371, 6.8\,AU, observed by HST \citep{rei97}.

\section{Discussion}

\subsection{The Binary Frequency of M6.0-M7.5 Stars}\label{bozomath}
One of the goals of our overall survey is to characterize the multiplicity of late M stars. \cite{clo04} report a binary fraction of $15\pm7\%$ for M8.0-L0.5 binaries with separations $>3$ AU. Complimenting their spectral type range, we present here the results from the largest flux limited (K$_s<12$) high-resolution imaging survey of M6.0-M7.5 primaries. From a sample of 30 target objects we have spatially resolved 3 systems that have companions. In order to estimate the binary fraction over this spectral range we examine two different techniques - Malmquist bias adjustment and flux ratio distribution limits.

We attempt to compensate for the leakage of equal magnitude binaries into our sample from further distances. The number of stars in the sample is adjusted upward by the ratio between the volume containing approximately 95\% of our detected binaries and the volume containing approximately 95\% of our target objects so as to compensate for fainter single stars that were not originally included. This results in a scaling factor of $(30.3/23.7)^3$\,=\,2.09. This gives a Malmquist corrected binary frequency of $3/(30\times2.09)=4.8\%$ with a Poisson uncertainty of $\pm2.8\%$. 

\cite{bur03} explored the limiting cases of the ratio of flux distributions emitted from binary stars for a magnitude-limited sample like ours. Using an observed uncorrected binary fraction of 0.1 (3/30), we consider 2 limiting cases for the flux ratio $f(\rho)$ distribution: 1) all the binary systems are of equal magnitude $(q=\rho=1)$ resulting in a binary fraction of $=4\%$; 2) the flux ratio distribution is flat and the binary fraction rises to $6\%$. This results in a binary frequency range of $4-6\%$ which is consistent with the $5\pm3\%$ we estimated when correcting for the Malmquist bias. \cite{bur03} derive a more accurate estimate of uncertainty for small samples using the binomial distribution. In this case the binary fraction is 5$_{-2}^{+4}\%$ (A. Burgasser, private communication). {\it Hence we conclude that for binaries with separations 3\,AU\,$<$\,a$<\,$300\,AU the M6.0-M7.5 binary frequency is within the range 5$_{-2}^{+4}\%$}.

This result is significantly lower than the $\sim32\%$ observed for earlier M0-M4 dwarfs \citep{fis92} but more comparable with the 15$\pm7\%$ for later M8.0/early L \citep{clo04}, 10\,-\,15$\%$ and 15$\pm5\%$ for L dwarfs \cite[respectively]{bou03,giz03} and 9$_{-4}^{+15}\%$ for T dwarfs \citep{bur03}. Our binary fraction is also comparable to open cluster surveys of M dwarfs in the Hyades \citep[11$\pm5\%$;][]{rei96} and brown dwarfs in the Pleiades \citep[15$_{-5}^{+15}\%$;][]{mar03}. We note the peculiar result of \cite{pin03} who predict binary frequencies of $\sim50\pm10\%$ for ultracool M dwarfs in the Pleiades and Praesepe open clusters using over-luminous positions on NIR CMDs to predict multiplicity. They claim the missing binaries have not been detected in the other surveys because the systems, \`a la PPl\,15 \citep{bas99}, are spectroscopic (a$_{proj}<1$\,AU) and beyond the resolution limit. However, \cite{rei02} conducted a spectroscopic binary survey of field M dwarfs and found 2 out of 36 targets. This implies a bias-corrected fraction of only 3$_{-2}^{+3}\%$ \citep{bur03}.

Is there a possible explanation for our lower result, 5$_{-2}^{+4}\%$, besides small number statistics? Over as narrow a spectral range as we have here, could there have been conditions in the local star formation process that led to a shortage of certain binary masses? We are more inclined to believe that these results are largely influenced by small number statistics and that our results are statistically consistent to within 1.5\,$\sigma$ of the later objects. {\it We conclude, therefore, that for binary systems with separations 3\,AU\,$<$\,a\,$<\,$300\,AU the M6.0-M7.5 binary frequency from our survey, while only within the range 5$_{-2}^{+4}\%$, is statistically consistent with less massive M, L, and T stars and significantly less common than that of G and early M stars.} 

\subsection{The Separation Distribution Function}\label{bozomath}
Even though our survey is sensitive out to 10$\arcsec$ ($\sim300$ AU), all 3 observed binary systems had projected separations $<0.3\arcsec$ (3\,-\,9\,AU) apart. We do not detect any wide ($>20$\,AU) VLM binary systems in our sample. Consequently, this rules out a sensitivity selection effect. This is consistent, and apparently without exception, in 31 other VLM binaries from several researchers as recorded in Table 4 of \cite{clo04} and 3 recent brown dwarf binary discoveries in the Pleiades \citep{mar03}. Sensitive to only 14\,AU, this may explain why \cite{rei96} found no Hyades brown dwarf binary systems from 53 HST PC observations. 

When we compare binary surveys that are sensitive to separations $\geq$\,2\,AU, there is consistency between the median separation of our 3 new binaries (5\,AU) and the peak distribution ($\sim$\,2-5 AU) of late M/early L binaries \citep{clo04}, L dwarfs \citep{giz03, bou03}, and T dwarfs \citep{bur03}. Going towards earlier spectral types there appears to be an increase in the semimajor axis separation. \cite{fis92} and \cite{duq91} show a broad separation peak of $\sim30$\ AU for early M and G dwarfs, respectively. {\it We conclude that the semimajor axes of M6.0-M7.5 binaries appear consistent with those of late M, L, and T dwarf systems but are significantly smaller on average than early M and G stars.} 

The difference between the separation distributions of these VLM binaries and more massive binary systems suggests that there may be important differences in their formation and evolution processes as well. This is further supported by the observed preference of equal-mass systems (q\,$\geq\,$0.7) as seen in this paper and for all the VLM binaries \citep{clo04,mar03} as compared to the flatter distributions of \cite{fis92} and \cite{duq91}. Ejection models \citep{rep01}, high-resolution hydrodynamical simulations \citep{dlg03}, collapsing turbulent molecular clouds \citep{bat02}, and multi-body decay models \citep{ste03} have not been able to predict both the $\sim$\,15$\%$ VLM binary fractions observed {\it and} the $\sim$\,60$\%$ binary fraction for the more massive $\sim$\,G-type stars. Our results reported here hope to add to the constraints of these and future formation and evolutionary models.

\section{Summary}\label{bozomath}

We have conducted the largest survey of nearby M6.0-M7.5 dwarfs with K$_s<12$ mag using the AO system Hokupa'a at the Gemini North telescope. The survey consisted of 30 stars and discovered 3 new binary systems with relatively equal mass components (q$\,>\,0.8$) and close projected separations of 0.12-0.29$\arcsec$ (3\,-\,10\,AU). While none of the binaries have been confirmed by common proper motions, they are very likely to be bound based on space density arguments. We have used various astrometric, observational, and statistical arguments to characterize the VLM binary frequencies and separations that contribute additional empirical constraints to binary formation mechanisms:
\begin{itemize}
\item We estimate the binary frequency of spectral type M6.0-M7.5 main sequence stars for separations 3\,AU\,$<$\,a$<\,$300\,AU from this survey to be 5$_{-2}^{+4}\%$. The figure is lower than, but statistically consistent with, slightly later type M, L, and T dwarfs to within 1.5\,$\sigma$. However, the frequency is less than that measured in studies of earlier M and G dwarfs.

\item The separations of the 3 systems are all $<10$ AU and consistent with the separations of later type M and L dwarfs (separation peak $\sim2$\,-\,5\,AU). This is in stark contrast with the broad peak separations of $\sim30$ AU for the more massive M and G binaries.

\item Two of the new systems are probable Hyades cluster members based on proper motions, moving cluster membership probabilities, radial velocities, and photometry. LP\,415-20 and LP\,475-855 have Hyades membership probabilities of 90$\%$ and 98$\%$, respectively. Our simulations show that only $\sim0.3\%$ of ultracool field M dwarfs at the photometric distances of the LP stars would have both Hyades membership probabilities $>$\,90$\%$ and radial velocities within 5\,km/s of the Hyades mean value. Both LP stars satisfy these criteria.

\item Triginometric distances would settle the question of membership. Assuming membership, the LP\,415-20 system is the tightest spatially resolved binary to date in the Hyades cluster with a projected separation of $3.6\pm0.7$ AU and the companions to LP\,415-20 and LP\,475-855 are among the lowest mass objects resolved to date in the cluster (0.081$_{-0.010}^{+0.009}$\,M$_{\sun}$ and 0.082$_{-0.009}^{+0.009}$\,M$_{\sun}$, respectively).

\end{itemize}

\acknowledgements
We thank the referee, Adam Burgasser, for many helpful comments ultimately leading to a significantly improved paper. We also thank Jim Liebert for helpful discussions regarding cluster membership. E. M. acknowledges support from the NASA Graduate Student Researchers Program (NGT5-50400). This research could not have been possible without support from AFOSR grant F49620-00-1-0294 and from NASA Origin of Solar Systems NAG5-12086. This paper is based on observations obtained with the Adaptive Optics System Hokupa`a/Quirc, developed and operated by the
University of Hawaii Adaptive Optics Group, with support from the National Science Foundation (NSF). These results were also based on observations obtained at the Gemini Observatory, which is operated by the Association of Universities
for Research in Astronomy, Inc., under a cooperative agreement with the NSF on behalf of
the Gemini partnership: the NSF (United States), the Particle
Physics and Astronomy Research Council (United Kingdom), the National Research
Council (Canada), CONICYT (Chile), the Australian Research Council (Australia), CNPq
(Brazil), and CONICET (Argentina). This
publication makes use of data products from the Two-Micron All-Sky Survey, which is a
joint project of the University of Massachusetts and the Infrared Processing and Analysis
Center/California Institute of Technology, funded by NASA and the NSF. Finally, the authors wish to recognize and acknowledge the very significant
cultural role and reverence that the summit of Mauna Kea has always had
within the indigenous Hawaiian community.  We are most fortunate and grateful to have
the opportunity to conduct observations from this mountain.

\clearpage
\begin{deluxetable}{llllllllll}
\tabletypesize{\scriptsize}
\tablecaption{M6.0-M7.5 Stars Observed with No Physical Companion Detections Between 0.1$\arcsec$-15$\arcsec$\label{tbl-1}}
\tablewidth{0pt}
\tablehead{
\colhead{2MASS Name} &
\colhead{Other Name} &
\colhead{K$_s$} &
\colhead{SpT} &
\colhead{Ref.} &
}
\startdata
2MASSI\,\,\,\, J0330050+240528 &LP 356-770& 11.36 & M7.0 & 1\\
2MASSI\,\,\,\, J0752239+161215 && \,\,\,9.82 & M7.0e & 2\\
2MASSI\,\,\,\, J0818580+233352 && 11.13 & M7.0 &  1\\
2MASSW J0952219-192431 && 10.85 & M7.0 & 1\\
2MASSW J1016347+275150 & LHS 2243 & 10.95 & M7.5 & 1\\
2MASSI\,\,\,\, J1024099+181553 && 11.21 & M7.0 & 1\\
2MASSW J1049414+253852 && 11.39 & M6.0 & 1\\
2MASSI\,\,\,\, J1124532+132253 && 10.03 & M6.5 & 2\\
2MASSW J1200329+204851 && 11.82 & M7.0 & 1\\
2MASSW J1237270-211748 && 11.64 & M6.0 & 1\\
2MASSW J1246517+314811 & LHS 2632 & 11.23 & M6.5 & 1\\
2MASSI\,\,\,\, J1253124+403404 && 11.20 & M7.5 & 4\\
2MASSW J1336504+475131 && 11.63 & M7.0 & 1\\
2MASSW J1344582+771551 && 11.83 & M7.0 & 1\\
2MASSI\,\,\,\, J1356414+434258 && 10.63 & M7.5 & 2\\
2MASSP\,\, J1524248+292535 && 10.15 & M7.5 & 3\\
2MASSW J1527194+413047 && 11.47 & M7.5 & 3\\
2MASSW J1543581+320642 &LP 328-36& 11.73 & M7.5 & 1\\
2MASSW J1546054+374946 & &11.42 & M7.5 & 1\\
2MASSW J1550381+304103 && 11.92 & M7.5 & 1\\
2MASSW J1757154+704201 &LP 44-162& 10.37 & M7.5 & 1\\
2MASSW J2052086-231809 &LP 872-22& 11.26 & M6.5 & 1\\
2MASSW J2221544+272907 && 11.52 & M6.0 & 1\\
2MASSW J2233478+354747 &LP 288-31& 11.88 & M6.0 & 1\\
2MASSI\,\,\,\, J2235490+184029 &LP 460-44& 11.33 & M7.0 & 1\\
2MASSW J2306292-050227 && 10.29 & M7.5 & 1\\
2MASSW J2313472+211729 &LP 461-11& 10.42 & M6.0 & 1\\
\enddata
\tablerefs{
(1) \cite{giz00} (2) Cruz et al. (in preparation) (3) \cite{rei02} (4) \cite{kir91}.
}
\end{deluxetable}

\clearpage
\begin{deluxetable}{lllllllll}
\tabletypesize{\scriptsize}
\tablecaption{The New Binary Systems \label{tbl-2}}
\tablewidth{0pt}
\tablehead{
\colhead{System} &
\colhead{$\Delta J$} &
\colhead{$\Delta H$} &
\colhead{$\Delta K^{\prime}$} &
\colhead{Sep. (mas)} &
\colhead{P.A. (deg)} &
\colhead{Date Observed (UT)} &
}
\startdata
LP\,415-20\tablenotemark{a} & $0.84\pm0.15$ & $0.77\pm0.10$ & $0.66\pm0.06$ & $119\pm8$ & $91.2\pm0.7$ & 2002 Feb. 07 \\
LP\,475-855\tablenotemark{b} & $0.48\pm0.05$ & $0.43\pm0.04$ & $0.48\pm0.03$ & $294\pm5$ & $131.6\pm0.5$ & 2001 Sep. 22 \\
2MASSW J1750129+442404 & $0.74\pm0.15$ & $0.73\pm0.15$ & $0.64\pm0.10$ & $158\pm5$ & $339.6\pm0.7$ & 2002 Apr. 25 \\
\enddata
\tablenotetext{a}{Also known as Bryja\,262.} 
\tablenotetext{b}{Also known as [LHD94]\,042614.2+13312 and 2MASSW\,J0429028+133759.}
\end{deluxetable}

\clearpage
\begin{deluxetable}{llllllllll}
\tabletypesize{\scriptsize}
\tablecaption{Summary of the New Binaries' A and B Components\label{tbl-3}}
\tablewidth{0pt}
\tablehead{
\colhead{Name} &
\colhead{$J$} &
\colhead{$H$} &
\colhead{$K_s$} &
\colhead{$M_{K_s}$\tablenotemark{a}} &
\colhead{SpT\tablenotemark{b}}&
\colhead{d$_{phot}$ (pc)\tablenotemark{c}} &
\colhead{Mass (M$_{\sun}$)\tablenotemark{d}} &
\colhead{Sep. (AU)} &
\colhead{P (yr)\tablenotemark{e}} 
}
\startdata
LP\,415-20A & $13.09\pm 0.06$ & $12.47\pm 0.05$ & $12.12\pm 0.04$ &\phn $9.72\pm 0.38$ & M7.0 & $30\pm5$ & $0.097_{-0.012}^{+0.011}$ & $3.6\pm 0.7$ & $23_{-6}^{+7}$ \\
LP\,415-20B &$13.93\pm 0.16$ &$13.24\pm 0.11$ &$12.78\pm 0.08$ & $10.37\pm0.39$&M9.5 & &$0.081_{-0.010}^{+0.009}$ \\

LP\,475-855A & $13.21\pm 0.04$ & $12.54\pm 0.04$ & $12.18\pm 0.04$ &\phn $9.84\pm0.36$ & M7.5 & $29\pm5$ & $0.093_{-0.009}^{+0.012}$ &  $8.6\pm 1.5$ & $86_{-19}^{+20}$ \\
LP\,475-855B&$13.69\pm 0.07$ & $12.97\pm 0.06$ &$12.66\pm 0.05$ & $10.32 \pm 0.36$&M9.5&& $0.082_{-0.009}^{+0.009}$  \\

2M1750A & $13.23\pm 0.06$ & $12.62\pm 0.06$ & $12.24\pm 0.05$ &\phn $9.77\pm0.39$ & M7.5 & $31\pm6$ & $0.097_{-0.012}^{+0.012}$ &  $4.9\pm 0.9$ & $36_{-9}^{+10}$ \\
2M1750B & $13.97\pm 0.16$ & $13.35\pm 0.16$ &$12.88\pm 0.11$ & $10.41 \pm 0.41$ & M9.5 && $0.085_{-0.016}^{+0.006}$ \\

\enddata
\tablenotetext{a}{M$_{K_s}$ = 7.65 + 2.13(J-$K_s$) with a rms $\sigma_{M_{Ks}}\,=\,0.33$. Relationship is valid for M6.5$<$SpT$<$L1. See \S 4.2 for more detail.}
\tablenotetext{b}{Spectral type estimated by SpT\,=\,3.54M$_{K_s}$\,-\,27.20 with $\pm1.5$ spectral subclasses of error in these estimates (SpT = 10 is defined as an\\ L0; valid for M6.5$<$SpT$<$L1). See \S 4.5 for more detail. }
\tablenotetext{c}{Distances based on M$_{K_s}$ as described in \S 4.2.} 
\tablenotetext{d}{Mass determination uses the models of \cite{cha00}. See Figures 7\,-\,9 in this paper. }
\tablenotetext{e}{Periods include a 1.26 multiplication of the projected separations compensating for random inclinations and eccentricities \citep{fis92}.}
\end{deluxetable}

\clearpage
\begin{deluxetable}{lccc}
\tabletypesize{\scriptsize}
\tablecaption{Hyades Members: A Comparison}
\tablehead{
\colhead{} &
\colhead{LP\,415-20} &
\colhead{LP\,475-855} &
\colhead{Hyades Cluster Center} 
}
\startdata
RA (J2000)   & 04:21:49.56\tablenotemark{a} & 04:29:02.83\tablenotemark{a}  & 04:26:47.96\tablenotemark{f}            \\
DEC (J2000) & +19:29:08.6\tablenotemark{a} & +13:37:59.2\tablenotemark{a}& +16:33:14.60\tablenotemark{f}       \\
proper motion ($\mu_\alpha,\mu_\delta$) (mas/yr) & (127,-36)\tablenotemark{a} & (103,-16)\tablenotemark{a} & $(106,-27)$\tablenotemark{i}\\
proper motion ($\mu_\|,\mu_\bot$) (mas/yr) & (132,-5)\tablenotemark{h} & (104,-2)\tablenotemark{h} & $(109,0)$\tablenotemark{b}\\
measured radial velocity (km/s)& 43.9$\pm$3.9\tablenotemark{d}, 36$\pm$4\tablenotemark{e}& 44.3$\pm2.0$\tablenotemark{c} & 39.5$\pm0.3$\tablenotemark{f}\\
predicted radial velocity (km/s) & 38.7$\pm0.3$\tablenotemark{b} & 40.1$\pm0.3$\tablenotemark{b} & 39.5$\pm0.3$\tablenotemark{f}  \\
photometric distance (pc) & $30.3\pm5.4$\tablenotemark{b} & $29.3\pm5.0$\tablenotemark{b} &  46.3$\pm0.3$\tablenotemark{g}\\ 
moving cluster distance (pc) & $41_{-4}^{+5}$\tablenotemark{h} & $47_{-6}^{+8}$\tablenotemark{h} &46.3$\pm0.3$\tablenotemark{g} \\ 
membership probability ($\%$) & 90\tablenotemark{h} & 98\tablenotemark{h} & 100 \\
\enddata
\tablerefs{
(a) \cite{giz00}; (b) calculated in this paper using the space motion and convergent point solution for the Hyades from \cite{deb01}; (c) \cite{rei02}; (d)\cite{sta94}; (e) \cite{jon96}; (f) \cite{deb01}; (g) 95$\%$ of Hyades members in \cite{per98} have distances to cluster center between 25-73\,pc; (h) calculated in this paper using the space motion from \cite{deb01} and proper motions from \cite{giz00}; (i) calculated in this paper using the space motion and distance from \cite{deb01}.    
}
\end{deluxetable}

\clearpage
\begin{deluxetable}{lclc}
\tabletypesize{\scriptsize}
\tablecaption{Estimated Temperatures \label{tbl-5}}
\tablewidth{0pt}
\tablehead{
\colhead{Object} &
\colhead{Spectral Type\tablenotemark{a}} &
\colhead{DUSTY (K)\tablenotemark{b}} &
\colhead{Dahn et al. (2002)/2MASS (K)\tablenotemark{c}} 
}
\startdata
LP\,415-20A & M7.0 & 2750$_{-170}^{+170}$ & 2600$\pm100$ \\
LP\,415-20B & M9.5 & 2460$_{-220}^{+190}$ & 2400$\pm100$ \\
LP\,475-855A& M7.5 & 2700$_{-170}^{+160}$ & 2600$\pm100$ \\
LP\,475-855B& M9.5 & 2480$_{-200}^{+170}$ & 2400$\pm100$ \\
2M1750A   & M7.5 & 2740$_{-190}^{+180}$ & 2600$\pm100$ \\
2M1750B   & M9.5 & 2460$_{-250}^{+180}$ & 2400$\pm100$ \\
\enddata
\tablenotetext{a}{Spectral types derived from M$_{K_s}$ vs. SpT fit in Figure \ref{fig10}.}
\tablenotetext{b}{Temperatures derived from M$_{K_s}$ and age estimates using DUSTY evolutionary tracks (Figures 7, 8, 9).}
\tablenotetext{c}{Temperatures derived from fit of field star temperatures covering spectral type range M6.5-L1.0 \cite[Table 5]{dah02}: T$_{eff}$\,=\,-113.5 SpT\,+\,3471.9  ($\sigma$\,=\,98.5\,K).} 
\end{deluxetable}

\clearpage

\clearpage

\clearpage

\clearpage

\clearpage

\clearpage

\clearpage
\begin{figure}
\includegraphics[angle=90,width=\columnwidth]{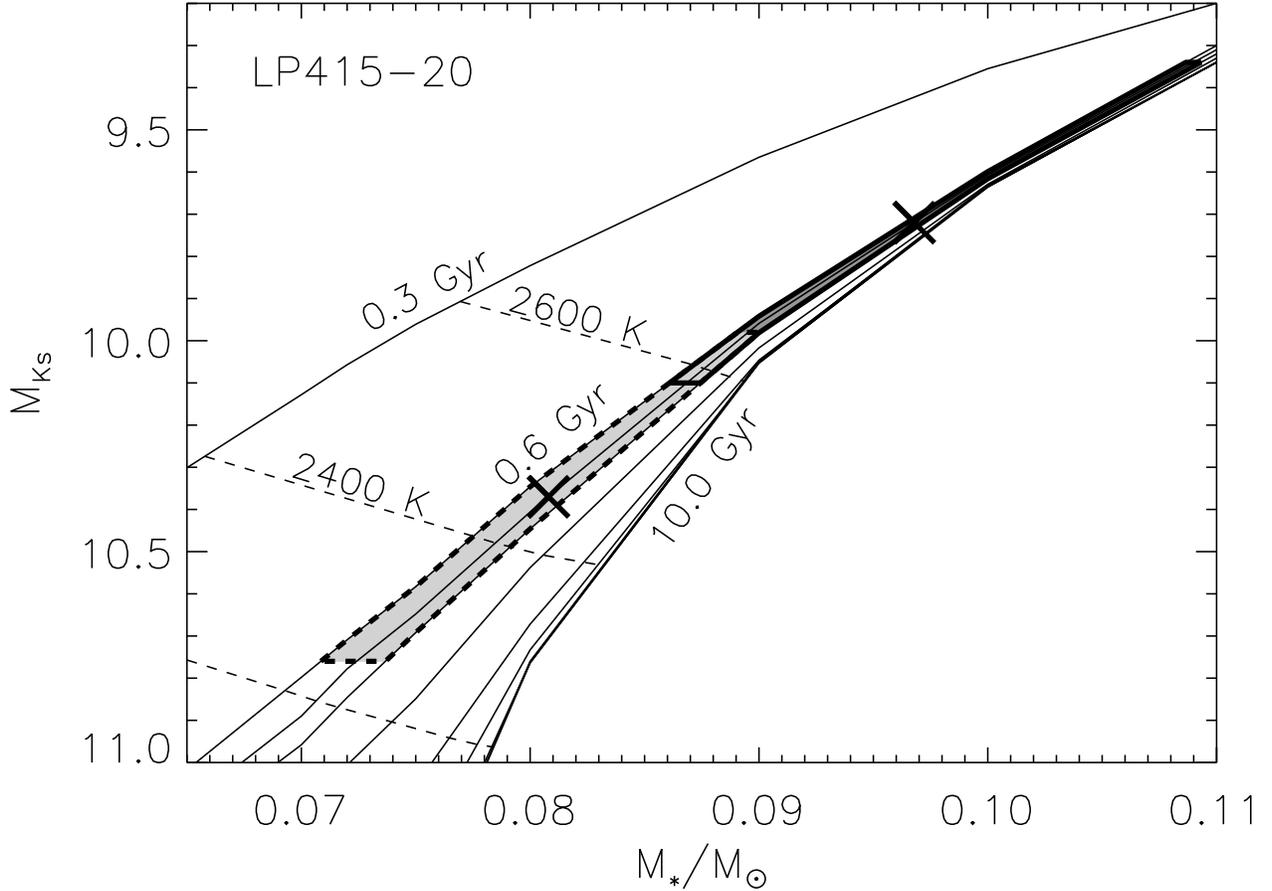}
\caption{\cite{cha00} DUSTY stellar and substellar evolutionary tracks custom integrated over the K$_s$ bandpass ([{\it m}/H]\,=\,0). The best-guess values of the 2 components of LP\,415-20 are indicated by the bold crosses with the primary at the top right and the companion at the bottom left. The polygons enclose the uncertainties in M$_{K_s}$ and age with the primary's outlined by a solid line and the companion's outlined by a dashed line. Absolute magnitude of the primary is derived from the Figure \ref{fig6} CMD; the observed $\Delta$K magnitude is added to obtain the secondary's absolute magnitude. Age is that of the Hyades (0.625$\pm0.050$\,Gyr). The model suggests a primary mass of 0.097$_{-0.011}^{+0.012}$\,M$_{\sun}$ and a temperature of 2750$_{-170}^{+170}$\,K. For the companion, the model predicts a mass of 0.081$_{-0.010}^{+0.009}$\,M$_{\sun}$ and a temperature of 2460$_{-220}^{+190}$\,K. The isochrones plotted are 0.3, 0.6, 0.65, 0.7, 0.85, 1.2, 1.7, 3.0, 5.0, 7.5, and 10.0 Gyr (the last 4 isochrones are indistinguishable at the given scaling). 
\label{fig7}} 
\end{figure}

\clearpage
\begin{figure}
\includegraphics[angle=90,width=\columnwidth]{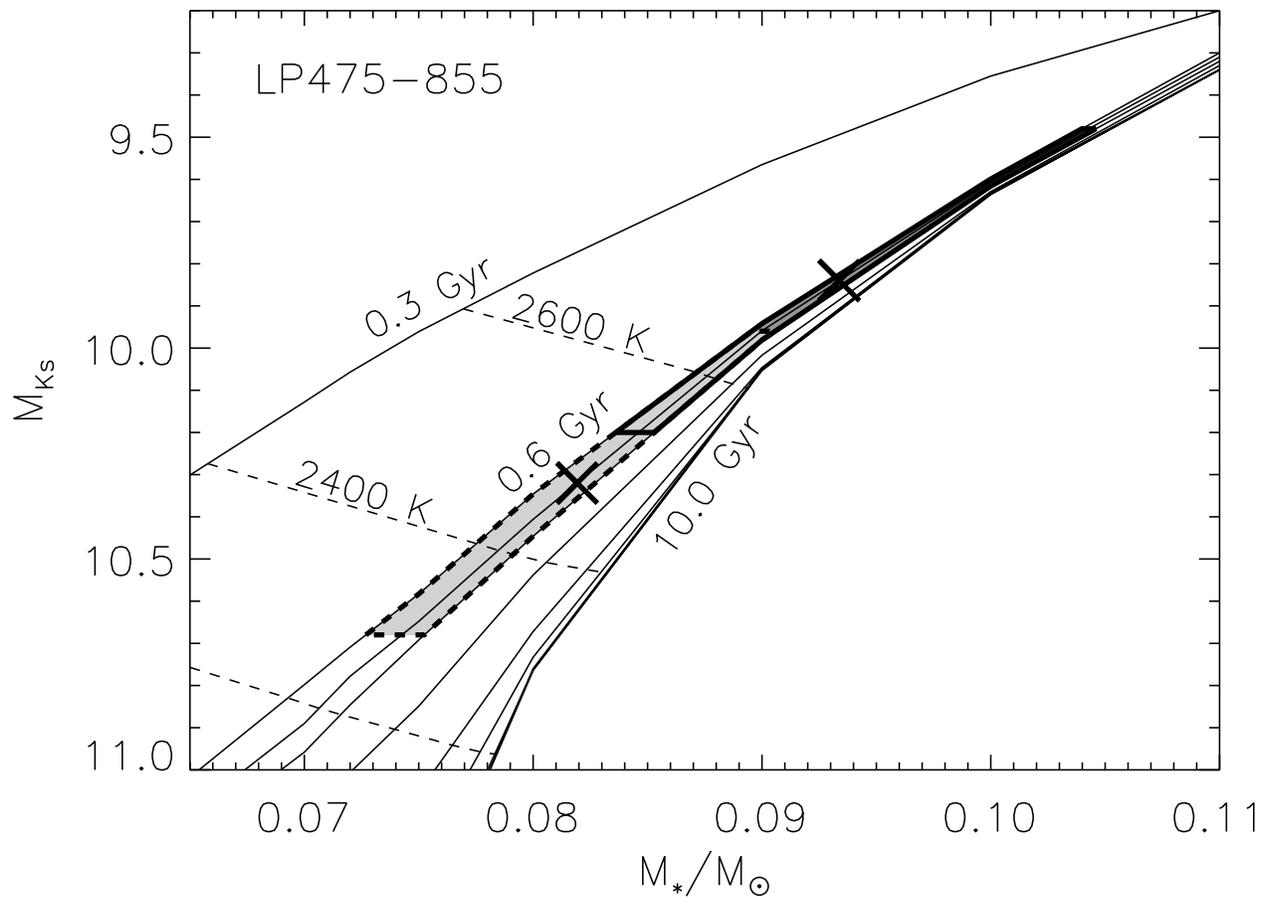}
\caption{As in Fig. 7 except for LP\,475-855. The model suggests a primary mass of 0.093$_{-0.009}^{+0.012}$\,M$_{\sun}$ and a temperature of 2700$_{-170}^{+160}$\,K. For the secondary the model suggests a mass of 0.082$_{-0.009}^{+0.009}$\,M$_{\sun}$ and temperature of 2480$_{-200}^{+170}$\,K.
\label{fig8}} 
\end{figure}

\clearpage
\begin{figure}
\includegraphics[angle=90,width=\columnwidth]{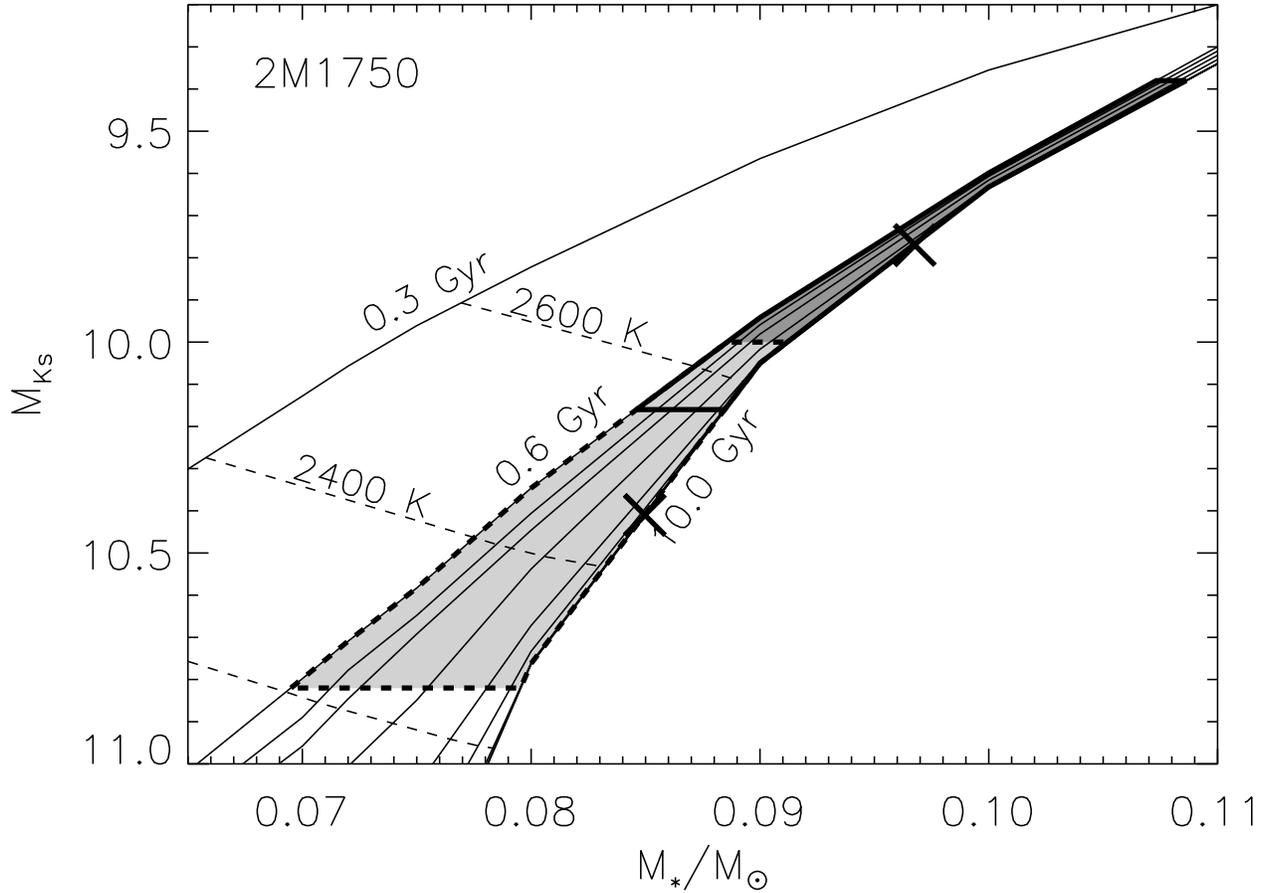}
\caption{As in Fig. 7 except for 2M1750 and a field age of $3.0_{-2.4}^{+4.5}$ for the system. The model suggests a primary mass of 0.097$_{-0.012}^{+0.012}$\,M$_{\sun}$ and a temperature of 2740$_{-190}^{+180}$\,K. For the secondary the model suggests a mass of 0.085$_{-0.016}^{+0.006}$\,M$_{\sun}$ and a temperature of 2460$_{-250}^{+180}$\,K.
\label{fig9}} 
\end{figure}

\end{document}